\documentstyle[12pt]{article}
\begin{document}
\thispagestyle{empty}
\begin{flushright}
{\bf TTP95-19\footnote{The complete postscript file of this
preprint, including figures, is available via anonymous ftp at
ttpux2.physik.uni-karlsruhe.de (129.13.102.139) as
/ttp95-19/ttp95-19.ps or via www at
http://ttpux2.physik.uni-karlsruhe.de/cgi-bin/preprints/
Report-no: TTP95-19}}\\
{hep-ph/9506256}\\
{\bf June 1995}\\
\end{flushright}
\vspace*{20mm}
\begin{center}
  \begin{large}
Electroweak Fermion-loop Contributions to
the Muon Anomalous Magnetic Moment\\
  \end{large}
  \vspace{0.5cm}
  \begin{large}
Andrzej Czarnecki and Bernd Krause
 \\
  \end{large}
  \vspace{0.1cm}
Institut f\"ur Theoretische Teilchenphysik, \\
Universit\"at Karlsruhe,
D-76128 Karlsruhe, Germany \\[4mm]
and \\[4mm]
  \begin{large}
William J.~Marciano
 \\
  \end{large}
  \vspace{0.1cm}
Physics Department\\
Brookhaven National Laboratory\\
Upton, New York 11973

\vspace{40mm}
  {\bf Abstract}\\
\vspace{0.3cm}
\noindent
\begin{minipage}{15.0cm}
The two-loop electroweak corrections to the anomalous magnetic moment
of the muon, generated by fermionic loops, are calculated.  An
interesting role of the top quark in the anomaly cancellation is
observed.  New corrections, including terms of order $G_\mu \alpha
m_t^2$, are computed and a class of diagrams previously thought to
vanish are found to be important.  The total fermionic correction is
$-(23\pm 3) \times 10^{-11}$ which decreases the electroweak effects
on $g-2$, predicted from one-loop calculations, by 12\%.  We give an
updated theoretical prediction for $g-2$ of the muon.
\end{minipage}
\end{center}
\newpage
The anomalous magnetic moment of the muon, $a_\mu \equiv (g_\mu-2)/2$,
provides a precision test of the standard model and potential window
to ``new physics'' effects.  The current experimental average
\cite{PDG}
\begin{eqnarray}
a_\mu^{\rm exp} = 116592300 (840) \times 10^{-11}
\end{eqnarray}
is in good accord with theory and constrains physics beyond the
standard model such as \cite{kno,km90} supersymmetry, excited leptons,
compositeness, etc.  A new experiment \cite{Hughes92} being prepared at
Brookhaven National Laboratory is expected to reduce the uncertainty
in  $a_\mu^{\rm exp}$ to below $\pm 40\times 10^{-11}$, more than a
factor of 20 improvement. At that level, electroweak loop corrections
become important and new physics at the multi-TeV scale is explored.

To fully exploit the anticipated experimental improvement, the standard
model theoretical prediction for $a_\mu$ should be known with
comparable precision.  That requires the confluence of calculational
effort involving very high order QED loops, hadronic loop
contributions, and even two loop electroweak effects.  Indeed, the
contributions to $a_\mu$ are traditionally divided up into
\begin{eqnarray}
a_\mu = a_\mu^{\rm QED} +a_\mu^{\rm Hadronic}+a_\mu^{\rm EW}
\end{eqnarray}
The QED loops have been computed to very high order \cite{r4}
\begin{eqnarray}
a_\mu^{\rm QED} &=& {\alpha\over 2 \pi}
+0.76585738(6) \left( {\alpha\over  \pi}\right)^2
+24.0454(4) \left( {\alpha\over  \pi}\right)^3
\nonumber\\ &&
+ 126.14(43) \left( {\alpha\over  \pi}\right)^4
+ 930(170) \left( {\alpha\over  \pi}\right)^5
\end{eqnarray}
where in the calculation of the $\tau$ lepton loops we used
$m_\tau=1777$ MeV.
Employing $\alpha^{-1} = 137.0359895(61)$ recommended by the PDG
\cite{PDG} gives
\begin{eqnarray}
a_\mu^{\rm QED} = 116584708(5) \times 10^{-11}
\end{eqnarray}
The uncertainty could be further reduced by a factor of 2, if we chose
to use $\alpha$ as determined from the electron $g_e-2$, $\alpha^{-1}
= 137.03599222(94)$; however, in either case it is well within the
$\pm 40\times 10^{-11}$ experimental goal.

A recent reexamination \cite{Jeg95} of hadronic vacuum polarization at
the ${\cal O}(\alpha/\pi)^2$ level, utilizing $e^+e^- \to {\rm hadrons}$ data
via a dispersion relation, gives
\begin{eqnarray}
a_\mu^{\rm Hadronic}({\rm vac.\,pol.}) = 7023.5(152.6)\times 10^{-11}
\label{eq:had}
\end{eqnarray}
Unfortunately, the uncertainty has not yet reached the hoped for level
of precision.  However, it is anticipated \cite{r6} that ongoing
improvements in $e^+e^- \to {\rm hadrons}$ data near the $\rho$ meson
resonance (which weighs heavily in (\ref{eq:had})) and theoretical
input in the higher energy region will significantly reduce the
uncertainty. Nevertheless, the goal of going below $\pm40\times
10^{-11}$ remains a formidable challenge.

The result in (\ref{eq:had}) must be supplemented by higher order,
${\cal O}(\alpha/\pi)^3$, hadronic vacuum polarization effects
\cite{kno,km90}
\begin{eqnarray}
a_\mu^{\rm Hadronic}({\rm H.O.\, vac.\,pol.})
=-90(5) \times 10^{-11}
\end{eqnarray}
and the light by light hadronic amplitudes \cite{bijn95,haya95}
\begin{eqnarray}
a_\mu^{\rm Hadronic}({\rm light\, by \, light})= 8(9)\times 10^{-11}\
\end{eqnarray}
Altogether, one finds
\begin{eqnarray}
a_\mu^{\rm Hadronic}= 6942(153)\times 10^{-11}
\end{eqnarray}

Now we come to the electroweak contributions to $a_\mu$, the primary
focus of this work.  At the one loop level, they are well known
\cite{fls72,Jackiw72,ACM72,Bars72,Bardeen72}
\begin{eqnarray}
\lefteqn{a_\mu^{\rm EW}(\rm 1\,loop) =
{5\over 3}{G_\mu m_\mu^2\over 8\sqrt{2}\pi^2}}
\nonumber\\ && \times
\left[1+{1\over 5}(1-4s_W^2)^2
+ {\cal O}\left({m_\mu^2 \over M^2}\right) \right]
\nonumber \\
&& = 195 \times 10^{-11}
\label{eq:oneloop}
\end{eqnarray}
where $G_\mu = 1.16639(1) \times 10^{-5}$ GeV$^{-2}$, $M=M_W$ or
$M_{\rm Higgs}$, and the weak mixing angle is defined by
$\sin^2\theta_W\equiv s_W^2 = 1-M_W^2/M_Z^2$.  We can safely
neglect the ${\cal O}\left({m_\mu^2 / M^2}\right)$ terms in
(\ref{eq:oneloop}). Also, throughout this paper we neglect terms suppressed
by the factor $1-4s_W^2$ whenever it simplifies the expressions without
affecting accuracy.

The one loop estimate of electroweak effects
is about 5 times the anticipated experimental accuracy.
Naively, one would expect higher order electroweak contributions to be
of relative ${\cal O}(\alpha/\pi)$ and hence insignificant.  However,
an interesting study by Kukhto, Kuraev, Schiller, and Silagadze
\cite{KKSS} (KKSS) found that not to be the case.  They showed that 2
loop electroweak contributions are quite large and must be included in
any serious estimate of $a_\mu^{\rm EW}$ or confrontation with future
experiments.

Including 2 loops and making the approximations mentioned above,
$a_\mu^{\rm EW}$ becomes
\begin{eqnarray}
a_\mu^{\rm EW} =
{5\over 3}{G_\mu m_\mu^2\over 8\sqrt{2}\pi^2}
\left( 1+ C{\alpha\over \pi} \right)
\end{eqnarray}
It is natural to separate the subset of the two loop electroweak
contributions which contain a fermion loop
\begin{eqnarray}
C= C^{\rm ferm} + C^{\rm bos}
\end{eqnarray}
Parts of both $C^{\rm ferm}$ and $C^{\rm bos}$
have been calculated by KKSS. Denoting the non calculated
contributions by $R_f$ (for fermionic loops) and by $R_b$ (for the
remaining diagrams) the KKSS results can be written as
\begin{eqnarray}
C^{\rm ferm} &=&  -{18\over 5}\ln\left( { M_Z^2\over m_\mu^2}\right)
-{9\over 5}\ln\left( { M_Z^2\over m_\tau^2}\right)
+1+{8\over 15}\pi^2 + R_f
\nonumber\\
C^{\rm bos} &=& -{49\over 15}\ln\left( { M_Z^2\over m_\mu^2}\right) +
R_b
\label{eq:KKSS}
\end{eqnarray}
The known parts  reduce
the electroweak contribution $a_\mu^{\rm EW}$ by about 24\% ($-46\times
10^{-11}$), a significant decrease.  A full calculation of $R_b$
is quite a daunting task
because of the large number of diagrams.  It has been estimated by
KKSS to influence the correction factor $C$ at the level of 10\%.
However, only a full 2 loop calculation will tell us if that is
the case.

In the present paper  we reexamine the fermionic loops contributing to
the two loop electroweak corrections and calculate $R_f$.
We find that a significant
subset of diagrams has been neglected in previous studies.  In
particular we find that the large logs of the ratios of $M_Z$ and
lepton masses contributing to $C^{\rm ferm}$ are cancelled by the
corresponding quark diagrams.
We also obtain new relatively large non-logarithmic corrections of
${\cal O}(G_\mu \alpha m_t^2)$ and ${\cal O}(1)$ terms.

In our calculation we chose the ratio of muon and vector boson masses
as an expansion parameter in the calculation of diagrams contributing
to $g-2$.  Such asymptotic expansions have recently obtained firm
theoretical foundation \cite{Smi94}.  After the expansion we still
have to perform two loop integrals, which however contain at most one
mass scale.  The calculation of such integrals is further facilitated
by the integration by parts method \cite{che81} and the symbolic
manipulation programs written in FORM \cite{form}.  In some cases we
used packages SHELL2 \cite{shell2} and MINCER
\cite{mincer} to check our results.

We use dimensional regularization with the dimension of space-time
equal $D=4-2\varepsilon$ and neglect terms containing $\gamma_E$ and $\ln
4\pi\mu$ which accompany the poles $1/\varepsilon$ and vanish in the sum;
this explains the appearance of logs of dimensionful quantities in the
intermediate (divergent) results.

For the discussion of hadronic loops we use the following quark masses
$m_u=m_d=0.3$ GeV, $m_s=0.5$ GeV, $m_c=1.5$ GeV, $m_b=4.5$ GeV,
$m_t=176$ GeV.  We perform the computations in the 't~Hooft-Feynman
gauge.  The basic two loop diagrams with fermion loop contributing to
muon $g-2$ are shown in fig.~1.  In addition we have to consider
diagrams obtained by replacing vector boson propagators by
corresponding Goldstone bosons as well as mirror counterparts of
asymmetric diagrams.

Contributions of diagrams with a fermion loop connected to the muon
line via two charged bosons are shown in fig.~1(a-c).
Isospin +1/2 fermions are denoted by $u$ and the isospin -1/2 fermions
by $d$.
We first consider the case when fermions in the loop  belong to
the first two generations.  Here the masses of the fermions in the
loop do not influence the result very much and we neglect them.
The ratio of the neglected terms to the result is at most of the order of
$
(m_c^2/ M_W^2)\ln (m_c^2/ M_W^2) < 0.3\%
$,
with $m_c$ denoting the mass of the charm quark $\approx 1.5$ GeV.
In ref.~\cite{KKS} it has been argued that the
diagrams 1(a) and 1(b) vanish by virtue of Furry's theorem.  We find
that this is not true even after adding contributions of all fermions
in a generation.  Furry's theorem consists in the observation that
contributions of diagrams with two different orientations of the
fermion loop mutually cancel.  This is not the case for diagrams 1(a)
and 1(b) because for  every fermion flavor interacting with the
external photon there is only one possible orientation of the fermion
line.  Only those parts of the expressions which contain a
single $\gamma_5$ cancel out after adding contributions from the
up-type quark, down-type quark and from the lepton.

Adding contributions of
quarks and leptons we obtain for a single light generation
(we put the relevant Kobayashi-Maskawa matrix elements equal 1)
\begin{eqnarray}
\Delta C^{\rm ferm}_{\rm light} = {2\over 3s_W^2}
\label{eq:light_sum}
\end{eqnarray}

In the third generation we can neglect only masses of $\tau$ and of
the $b$ quark.
For the $\tau$ lepton loop we obtain
\begin{eqnarray}
\Delta C^{\rm ferm}_\tau = {1 \over 60s_W^2}
\label{eq:tau}
\end{eqnarray}

For the sum of all diagrams containing top and bottom quark loops we
find
\begin{eqnarray}
\Delta C^{\rm ferm}_{\rm tb} &= &
        {3\over 5s_W^2}   \left[
         {m_t^2 \over M_W^2}  \left(
          - {8 \over 3}
          - {5 \over 4 \varepsilon}
          + {5 \over 2} \ln (m_tM_W)
          \right)
\right. \nonumber \\ && \left.
-{1\over 12} -{1\over 2} \ln {m_t^2\over M_W^2}
\right]
\label{eq:charged}
\end{eqnarray}
We computed  ${\cal O}(M_W^2/m_t^2)$ and ${\cal O}(M_W^4/m_t^4)$
corrections to this formula. Those terms turn out to have small
coefficients which render them  numerically insignificant.
The singular terms $m_t^2/\varepsilon$ will be canceled by renormalization of
the $W$ boson mass present in the one loop electroweak contributions
to muon $g-2$.

We now consider the diagrams with a photon-photon-$Z$ coupling
induced by fermion loops shown in fig.~1(d).  For electron or muon in
the loop we obtain
\begin{eqnarray}
\Delta C^{\rm ferm}_{1d}(e)   &=&
   - {9\over 5}\left( \ln{M_Z^2 \over m^2_\mu}+{5\over 6} \right)
\nonumber\\
\Delta C^{\rm ferm}_{1d}(\mu)   &=&
 - {9\over 5} \left( \ln{M_Z^2 \over m^2_\mu}
 - {8\over 27}\pi^2+ {11\over 18}
 \right)
\end{eqnarray}
in agreement with \cite{KKSS}.  In that reference one also finds a formula
for the $\tau$ lepton which can be generalized  for all fermions
sufficiently heavier than the muon and lighter than the $Z$ boson
\begin{eqnarray}
\Delta C^{\rm ferm}_{1d}(f)   &=& {18\over 5}I_{3f}Q_f^2
\left( \ln{M_Z^2 \over m^2_f}-2 \right)
\label{eq:logsMZ}
\end{eqnarray}
In practice this formula can be used for all quark loops except for
the top quark, for which we find
\begin{eqnarray}
\Delta C^{\rm ferm}_{1d}(t)   &=& {M_Z^2\over m_t^2}
\left(
 {2 \over 3} + {2 \over 5} \ln{m_t^2\over M_Z^2}
\right)
\end{eqnarray}
In ref.~\cite{KKSS} the total fermionic two loop effect on muon
$g-2$ was estimated by summing only electron, muon, and $\tau$
contributions to diagram 1(d).  It has been concluded that the source
of large corrections are logarithms of ratios of these light fermion
masses to the mass of the $Z$ boson. Such treatment is incomplete
and misleading.   We can
see from the formula (\ref{eq:logsMZ}) that the sum over {\em all} fermions
(see discussion in \cite{km90})
in the first two generations leads to the cancellation of $M_Z$
dependent logarithms, due to the no-anomaly condition
$
\sum_f I_{3f}Q_f^2 =0
$.
This pattern no longer holds for the third generation; here, due to
the large mass of the top quark, its contribution is suppressed by a
factor $M_Z^2/m_t^2$.  This leads to the appearance of the
logarithm of $Z$ boson mass in the sum of all contributions to 1(d)
\begin{eqnarray}
\Delta C^{\rm ferm}_{1d} &=& -{18\over 5}\ln { (m_u m_c M_Z)^{4/3}\over
(m_d m_s m_b)^{1/3}  m_\mu^2 m_\tau}
\nonumber\\ &&\!\!
-
{5} +{8\over 15}\pi^2
+ {M_Z^2\over m_t^2}
\left(
{2 \over 3}  +  {2 \over 5} \ln{m_t^2\over M_Z^2}
\right)
\label{eq:2a}
\end{eqnarray}
The first line of (\ref{eq:2a}) gives the dominant contributions of
the diagram 1(d) from all fermions.  We note that the mass of the top
quark is absent in this part reflecting the suppression of top loop
discussed above.  The second line summarizes the corrections to the
dominant effect from the electron, muon,
and from the top quark.  Numerically we obtain
$\Delta a_{1d} = -14.4 \times 10^{-11}$ in contrast with the value
given in \cite{KKS,KKSS} $\Delta a_{1d(e,\mu,\tau)} \approx -25.6
\times 10^{-11}$.  This reduction of the correction is caused by the
cancellation of the $M_Z$ dependent logarithms.  We stress that the
$M_Z$ which is still present in the main part of (\ref{eq:2a}) is
caused by the large mass of the top quark and suppression of its
contribution.

To summarize this part we note that the large numerical value of
the sum of diagrams 1(d) is generated by large mass splittings among
the fermions.  This is the main difference between our result and the
result of \cite{KKSS}: imagine a model in which all fermions had equal
masses, then the sum of the three leptonic contributions discussed in
\cite{KKSS} would be equal $3\times \Delta a_{1d(\mu)}$, whereas we
find that the total correction due to 1(d) would vanish.  We believe
to have found a qualitatively new type of the top quark effect: it is
namely the extremely large mass of the top quark which determines the
shape of the main part of the formula (\ref{eq:2a}), although the
numerical value of $m_t$ is completely irrelevant.  What is important
is that $M_Z^2/m_t^2 \ll 1$.

We now discuss the remaining diagrams, neglected
in \cite{KKSS}.
In the diagram 1(e) we have to distinguish two cases, again treating
top quark separately.
For the light fermions we find, in contrast to the diagram 1(d), that
the logarithmic factors are suppressed by
extra powers of $m_f^2/M_Z^2$; we retain good accuracy by taking
massless fermions.  We find
\begin{eqnarray}
\Delta C^{\rm ferm}_{1e}({\rm light}\, f)   = -
{I_{f3}^2 - 2s_W^2 I_{f3}Q_f + 2 s_W^4 Q_f^2 \over 15 s_W^2 c_W^2}
\end{eqnarray}
This contribution becomes sizable after adding
the top quark effect.  For all fermions together we find
\begin{eqnarray}
\lefteqn{\Delta C^{\rm ferm}_{1e}  = -{3 \over 5s_W^2c_W^2}
\left[
         {2 \over 3}
       - {4 \over 3}  s_W^2
       + {16 \over 9}  s_W^4
\right.}
\nonumber\\ &&
\left. - {m_t^2\over M_Z^2}
 \left( {17 \over 24}
  + {5 \over 8\varepsilon} -
  {3 \over 8}\ln m_\mu^2 - {5 \over 8}\ln m_t^2
          - {1 \over 4}\ln M_Z^2 \right)
\right]\!\!
\label{eq:2b}
\end{eqnarray}
where we neglected terms ${\cal O}(M_Z^2/m_t^2)$ which proved to be
small.

In the remaining diagrams we have a scalar particle coupling to the
fermion loop, and therefore we only consider top loops.  If the $Z$
boson in the diagram 1(d) is replaced by the neutral Goldstone boson
we obtain
\begin{eqnarray}
\Delta C^{\rm ferm}_{1d(G)}(t)   &=&
       - {16 \over 5} - {8 \over 5}\ln {m_t^2 \over M_Z^2}
\label{eq:2c}
\end{eqnarray}
Finally, there is the diagram 1(f)  containing the Higgs boson.
It is the only non-negligible contribution of the Higgs
boson among two loop fermion diagrams.
We consider
three cases, depending on the hierarchy of masses of the top quark and
the Higgs.  For $M_H \ll m_t$ we get
\begin{eqnarray}
\lefteqn{\Delta C^{\rm ferm}_{1f}(t)   =
       - {104 \over 45} - {16 \over 15}\ln {m_t^2 \over M_H^2} }
\nonumber \\&&
\to \Delta a_\mu^H = -2.1\times 10^{-11}\quad (\mbox{for }M_H=60 \mbox{ GeV})
\label{eq:lightH}
\end{eqnarray}
For the case of $M_H \gg m_t$ we observe stronger
suppression of this amplitude
\begin{eqnarray}
\lefteqn{\Delta C^{\rm ferm}_{1f}(t)   = -{m_t^2 \over M_H^2}
\left[
{24\over 5}
+{8\over 15}\pi^2
+{8\over 5}
\left(
\ln {M_H^2 \over m_t^2}
-1
\right)^2
\right]}
\nonumber \\ &&
\to \Delta a_\mu^H =
-1.6\times 10^{-11}\quad (\mbox{for }M_H=300 \mbox{ GeV})
\label{eq:heavyH}
\end{eqnarray}
In order to estimate the size of the Higgs diagram in the case of
similar top and Higgs masses we put $m_t = M_H$.  In this case we find
\begin{eqnarray}
\Delta C^{\rm ferm}_{1f}(t)   &=& -{32\over 5}
\left[
1-{1\over \sqrt{3}}{\rm Cl}_2\left( {\pi\over 3} \right)
\right]\qquad (M_H=m_t)
\nonumber \\&&
\to \Delta a_\mu^H = -1.2\times 10^{-11}
\label{eq:equalHt}
\end{eqnarray}
where ${\rm Cl}_2$ is the Clausen function \cite{lewin}.
The contribution of the Higgs boson is small and we approximate it by
$-1.5(\pm 1.0)\times 10^{-11}$.

A few words are in order to explain why several diagrams with neutral
bosons have been omitted in fig.~1.  Diagrams with two scalar bosons
($HH$, $HG^0$, $G^0G^0$) coupling to the muon line are at most of the
order $m_\mu^4/M_Z^4$.  The remaining diagrams (e.g.~$ZH$) are either
exactly zero, or are suppressed by the vector coupling of the $Z$
boson to the muon (factor $1-4s_W^2$).

Some large two loop corrections can be absorbed in the one loop result
if it is parametrized in terms of
$G_\mu$ determined from the muon's life time.  This corresponds to the
replacement of bare parameters
\begin{eqnarray}
{e^2\over 8 s_W^2 M_W^2}\to {G_\mu \over \sqrt{2}} (1-\Delta)
\end{eqnarray}
where $\Delta$ is determined by studying electroweak corrections to
the muon decay
width.
Because we are interested in fermion loops, only quark and lepton
corrections to the W propagator need be included.  They induce the
following counterterm which cancels the divergences we encountered in
the charged boson diagrams (eq.~\ref{eq:charged}) and the $Z$ boson
vacuum polarization (eq.~\ref{eq:2b})
\begin{eqnarray}
\lefteqn{\Delta C^{\rm ferm}_{c.t.}= -{2\over s_W^2}
+{2\over 5s_W^2c_W^2}
\left(
1 - 2s_W^2+{8\over 3}s_W^4
\right)}
 \nonumber\\ &&
-{3\over 4s_W^2}{m_t^2\over M_W^2}\left( -{1\over 2\varepsilon}
+{1\over 2}\ln m_t^2
-{1\over 5}\ln M_Z^2
+          \ln M_W^2
-{3\over 10}\ln m_\mu^2
-{79\over 60} \right)
\label{eq:ct}
\end{eqnarray}

The final result for the fermion loop effect on muon $g-2$ is
obtained by adding the contributions given by equations
(\ref{eq:light_sum},\ref{eq:tau},\ref{eq:charged},\ref{eq:2a},\hspace{-2mm}
\ref{eq:2b},\ref{eq:2c},\ref{eq:ct}) and the contribution of the Higgs
boson diagram $\Delta C^{\rm ferm}_{1f}(t)$
given by approximating eq.~(\ref{eq:lightH},
\ref{eq:heavyH},\ref{eq:equalHt}).  Our final formula is
\begin{eqnarray}
C^{\rm ferm} &=&
-{18\over 5}
\ln{ (m_u m_c M_Z)^{4/3} \over (m_d m_s m_b)^{1/3} m_\mu^2 m_\tau }
\nonumber\\
&&
-{3 \over 16}{m_t^2 \over  s_W^2 M_W^2}
- {3\over 10s_W^2}\ln{m_t^2\over M_W^2}
\nonumber\\
&&
- {8\over 5} \ln{m_t^2\over M_Z^2}
- {41\over 5} - {7\over 10s_W^2}
+ {8 \over 15} \pi^2 + \Delta C^{\rm ferm}_{1f}(t).
\label{eq:fin}
\end{eqnarray}
This is our main result which replaces the old estimate of $C^{\rm
ferm}$ given in (\ref{eq:KKSS}).
We dropped all terms
suppressed by negative powers of $m_t$.  We checked explicitly that
their numerical impact is negligible.

The ${\cal O}(m_t^2/M_W^2)$ term in eq.~(\ref{eq:fin}) is related to
the $\rho$ parameter that appears in the ratio of weak neutral to
charged current amplitudes and comparisons of the $W^\pm$ and $Z$
masses.  It can be viewed as an induced correction brought about by
our renormalization of the one loop $Z$ contribution in terms of
$G_\mu$, a charged current parameter.  We also note that except for their
incomplete cancellation in the anomaly diagrams of fig.~1(d), no other effect
of the 2 light fermion generations resides in our final result.

For the numerical evaluation of the remaining terms which contain the
weak mixing angle we take $s_W^2=0.223$.
We obtain
\begin{eqnarray}
C^{\rm ferm} = -50(6)
\end{eqnarray}
which means that
the correction to muon anomalous magnetic moment $a_\mu$
from the fermion loops is $-(23\pm 3) \times 10^{-11}$.
The theoretical uncertainty has several sources: the unknown
mass of the Higgs boson, uncertainty in the masses of the light quarks
which parametrize the hadronic effects, and the large experimental
error in the present value of $m_t$.  Finally, higher order three loop
contributions remain unknown.  Altogether we estimate these effects to
yield an uncertainty at the level of $3\times 10^{-11}$, more than an
order of magnitude below the predicted experimental precision.

Including the fermionic two loop corrections and partial two loop
bosonic effects, we obtain the updated theoretical predictions
\begin{eqnarray}
a_\mu^{\rm EW} &=& (152(3) + 0.45R_b)\times 10^{-11}
\nonumber\\
a_\mu^{\rm th} &=& (116591802(153)+0.45R_b)\times 10^{-11}
\end{eqnarray}
What remains is to compute $R_b$ and lower the hadronic loop
uncertainty.  Work on both is in progress.

After completing this calculation we learned about ref.~\cite{Peris95}
which contains an analysis of quark contributions to the diagram in
fig.~1(d).  For the light quarks their numerical result obtained using
the chiral perturbation theory is the same as our evaluation using
constituent quark masses.  The large difference between their final
evaluation of the fermionic loops and our result is due
to diagrams of fig.~1(a-c,e,f) which were not considered in
\cite{Peris95}.

\section*{Acknowledgement}
It is a pleasure to thank M.~Steinhauser for collaboration at the
early stage of this project.  A.C. and B.K. also thank K.G.~Chetyrkin
for many helpful discussions and J.H.~K\"uhn for interest in this work
and support.
This work was supported by BMFT 056 KA 93P and by ``Graduiertenkolleg
Elementarteilchenphysik'' at the Karlsruhe University.

\end{document}